# Unexpectedly super strong paramagnetism of aromatic peptides due to cations of divalent metals

Haijun Yang[1,2,#], Liuhua Mu[2,9,#], Lei Zhang[3,#], Zixin Wang[4], Shiqi Sheng[5], Yongshun Song[5], Peng Xiu[6], Jihong Wang[5], Guosheng Shi[7], Jun Hu[1,2], Xin Zhang[3*], Feng Zhang[4,8*], Haiping Fang[1,5*]

1. Shanghai Synchrotron Radiation Facility, Zhangjiang Laboratory (SSRF, ZJLab), Shanghai Advanced Research Institute, Chinese Academy of Sciences, Shanghai 201204, China
2. Shanghai Institute of Applied Physics, Chinese Academy of Sciences, Shanghai 201800, China
3. High Magnetic Field Laboratory, Key Laboratory of High Magnetic Field and Ion Beam Physical Biology, Hefei Institutes of Physical Science, Chinese Academy of Sciences, Hefei 230031, China
4. State Key Laboratory of Respiratory Disease, Guangzhou Institute of Oral Disease, Stomatology Hospital, Department of Biomedical Engineering, School of Basic Medical Sciences, Guangzhou Medical University, Guangzhou 511436, China
5. School of Science, East China University of Science and Technology, Shanghai 200237, China
6. Department of Engineering Mechanics, Zhejiang University, Hangzhou 310027, China
7. Shanghai Applied Radiation Institute, Shanghai University, Shanghai 200444, China
8. Wenzhou Institute, University of Chinese Academy of Sciences, 16 Xinsan Road, Wenzhou 325001, China
9. University of Chinese Academy of Sciences, Beijing 100049, China

[#]These authors contributed equally to this work.

*Corresponding author. Email: fanghaiping@sinap.ac.cn; fengzhang1978@hotmail.com; xinzhang@hmfl.ac.cn

Abstract:
**The magnetism of most biological systems has not been characterized, which directly impedes our understanding of many magnetic field-related phenomena, including magnetoreception and magnetic bio-effects. Here we measured the magnetic susceptibility of aromatic peptide AYFFF self-assemblies in the presence or absence of divalent metal cations in liquid phase at room temperature. Unexpectedly, the magnetic susceptibilities of AYFFF self-assemblies in the chloride solution of various divalent cations ($Mg^{2+}$, $Zn^{2+}$, and $Cu^{2+}$) show super strong paramagnetism. We attribute the super strong paramagnetism to existence of the magnetic moments on the cations adsorbed on aromatic rings in the AYFFF assemblies through hydrated cation–π interactions, where the adsorbed cations**

**display non-divalent behavior with unpaired electron spins. Our results indicate the super strong paramagnetism or potential ferromagnetism in the aromatic ring-enriched biomolecules when there are enough cations of divalent metals adsorbed. The findings not only provide fundamental information for understanding the magnetism of biological systems, provoke insights for investigating the origin of magnetoreception and bio-effects of magnetic fields, but also help developing future magnetic-control techniques on aromatic ring-enriched biomolecules and drugs in living organisms, as well as biomaterial fabrication and manipulation.**

**Introduction**

It has been known for centuries that a large passel of animals navigates using Earth's magnetic field. Later, scientists found that microtubule and mitotic spindle orientation, DNA synthesis, cell cycle, proliferation, orientation, and zygote division can be significantly affected by high static magnetic fields (1-5). In addition, it was found that self-assembled aromatic-peptide nanotubes can align in high magnetic fields, which mainly originates from the well-ordered stacking of aromatic rings in the peptide nanotube (6, 7), given that the aromatic rings have a large diamagnetic anisotropy (8, 9).

Although the effects of high magnetic field on some biological samples have been clearly demonstrated (10-13), it is still unclear that how living organisms reliably sense weak magnetic field, such as the geomagnetic field (~50 μT), in the presence of thermal fluctuations and other sources of noise. Whether the magnetoreception in animals originates from magnetic nanostructures is still under considerable debate. In 2015, Qin et al. reported a new putative magnetic receptor (MagR) capable of sensing the geomagnetic field (14); however, this finding subsequently was challenged by Meister, who argued that MagR contains only 40 Fe atoms spreading out over a generous 24 nm, whose interaction with the geomagnetic field is too small (by at least five orders of magnitude) to overcome thermal fluctuations at room temperature (15). Moreover, it was proposed that the iron-containing sensory dendrites in the upper beak of pigeons are responsible for magnetoreception (16), but subsequent works argued that they are clusters of iron-rich macrophages (17, 18). Meanwhile, there have been efforts trying to understand the magnetism of biological components without ferromagnetic components, such as the radical-pair mechanism with cryptochromes (19, 20). However, there is still no fully satisfactory answer, as mentioned in a recent review paper, "Does Quantum biology exist (that is, is magnetic sensing truly quantum in at least some animals)?" (21) Certainly, we may have another fundamentally important question: Do unconventional magnetosensing matters (i.e., magnetic-particle-independent) exist in living organisms so that they can interact strongly with weak magnetic fields?

Very recently, we demonstrated that calcium ions on graphene showed ferromagnetic property (22), despite the traditional non-ferromagnetic nature of calcium ions. The key to this unexpected ferromagnetic property lies in the existence of magnetic moments on the Ca cations adsorbed on aromatic rings of graphene through strong cation–π interactions, where the adsorbed cations display monovalent behavior with unpaired electron spins and the two dimensional crystal is in the form of CaCl of unconventional stoichiometries, as the two-dimensional $Na_2Cl$ observed earlier (23). It is noteworthy that the cation–π interaction, a kind of non-covalent interaction, has attracted much attention (22-29), and the discovery of the above-mentioned crystals of unconventional stoichiometries is benefited from the polycyclic aromatic rings with delocalized π bonds on graphene, which greatly enhance the cation–π interactions. Along this direction, we have already theoretically predicted and experimentally demonstrated the cationic controlled graphene oxide membranes for ion sieving (27) and high salt accumulation inside carbon nanotubes soaked in dilute solutions (28). However, for biomolecules rich in aromatic rings which are not covalently connected with each other, it is reasonable to expect that their cation-π interactions are relatively weak. Therefore, when adsorbed onto such aromatic rings in biomolecules, can the divalent ions originally in the solution still exhibit the non-divalent behaviour with the structure similar to CaCl (with only one chloride ion per calcium on average) on graphene? Do they also display strong paramagnetic or even ferromagnetic properties?

To answer this question, in this paper, we take the AYFFF aromatic peptide as an example, while using the IIIGK non-aromatic peptide as a negative control. We found that AYFFF assemblies dispersed in the $MgCl_2$, $ZnCl_2$, and $CuCl_2$ solutions of sufficiently high concentrations displayed super strong paramagnetism, which might even approach the mass susceptibility of ferromagnetism, at room temperature. In contrast, the assemblies of IIIGK and AYFFF peptides in pure water displayed diamagnetic properties in all conditions. Therefore, we attribute the super strong paramagnetism to the cations of divalent metals. Further theoretical and experimental studies show that the key to the observed super strong paramagnetism lies in the adsorption of cations on the aromatic rings via cation-π interactions, in which the cations display non-divalent behavior and thus possess magnetic moments.

**Results and discussion**

In our experiment, the AYFFF aromatic and IIIGK non-aromatic peptide powders were first dispersed into pure water (Milli-Q, 18.2 MΩ) with the concentration of 0.5 mg/mL, and stored still at 20 °C for 3 days to form peptide assemblies. The dispersions were then thoroughly mixed with MgCl$_2$ solutions to obtain mixtures with different MgCl$_2$ concentrations. After settling for 30 minutes, the supernatant was taken out for morphology characterization and magnetic susceptibility measurement. Figure 1a displays the typical self-assembled AYFFF peptide fibers in the supernatant, which are longer than 2.0 μm with the height of ~8.9 nm.

We used a Quantum Design MPMS3 SQUID magnetometer to measure the direct current (DC) magnetic susceptibility at 293 K (20°C). Peptide mixtures of about 160 μL were loaded into a liquid sample holder (C130D, Quantum Design), which was detected for the leakproofness, and sealed tightly before experiment. The magnetic field was swept between -30,000 Oe and 30,000 Oe with one measurement point per 2500 Oe. All DC magnetic susceptibilities were corrected for magnetic contribution from the sample holder, and the solvent (pure water or salt solution) by directly subtracting their weight-scaled voltage signals and then fitting with a SquidLab program (30). As a result, we can get the magnetization ($M$) versus magnetic field ($H$) curves, and compute the mass susceptibility ($\chi$), $\chi = M/H = a/mH$, where $a$ is the moment measured by the SQUID magnetometer and $m$ is the mass of the assembled peptide. We chose cgs units in this study.

Figure 1b displays a typical example of the magnetization ($M$) measured with respect to the magnetic field ($H$) applied for the AYFFF peptide assemblies in 40 mM MgCl$_2$ solution. We linearly fit the magnetization ($M$) with respect to the magnetic field ($H$). The slop is $1.20 \times 10^{-4}$ emu/g, which is three orders of magnitude larger than the absolute value of the susceptibility of water. In Fig. 1c, we show the mass susceptibility ($\chi$) for the peptide in at least three independently repeated trials. The $\chi$ ranges from $4.90 \times 10^{-5}$ emu/g to $3.84 \times 10^{-4}$ emu/g, with the average value of $1.50 \times 10^{-4}$ emu/g in 40 mM MgCl$_2$ solution, showing super strong paramagnetism of these peptide assemblies in salt solutions. The relatively large fluctuation of $\chi$ is likely due to the morphology variations of the AYFFF assemblies. Moreover, the average values of the $\chi$ are $2.08 \times 10^{-4}$ emu/g, $9.68 \times 10^{-5}$ emu/g, $2.05 \times 10^{-4}$ emu/g, $1.28 \times 10^{-4}$ emu/g, and $2.83 \times 10^{-4}$ emu/g for the AYFFF peptide assemblies in the ZnCl$_2$ and CuCl$_2$ solutions of 40 mM, and in the MgCl$_2$, ZnCl$_2$, and CuCl$_2$ solutions of 20 mM, respectively, suggesting that the super strong paramagnetism of the assembled nanofibers of AYFFF in the chloride solution of divalent cations should be universal. It is worth emphasizing that the salt is of critical importance for the observed super strong paramagnetism. Our further experiment shows that the AYFFF assemblies in pure water without any salt have a negative average susceptibility of $-3.14 \times 10^{-5}$ emu/g, consistent with the strong diamagnetism observed in our previous work (31).

Since paramagnetic and diamagnetic objects are subjected to magnetic forces of opposite directions in a given gradient magnetic field, we also measured $\chi$ for the peptide assemblies by weighing the magnetic force of peptides at different states under a fixed magnetic field by using a high-precision electronic balance (XPR205, METTLER TOLEDO, Switzerland) as described in Ref. (31) (Fig. 1c, blue dots). Our results show that the χ of AYFFF peptide assemblies in 40 mM $MgCl_2$ solution ranges from $6.12 \times 10^{-5}$ emu/g to $4.07 \times 10^{-4}$ emu/g with an average value of $1.90 \times 10^{-4}$ emu/g, which is close to the data obtained by the SQUID magnetometer, $1.50 \times 10^{-4}$ emu/g. We have further measured χ for the peptide in chloride solutions of $ZnCl_2$ and $CuCl_2$ with same concentrations and got similar results as by SQUID measurement (Fig. 1c).

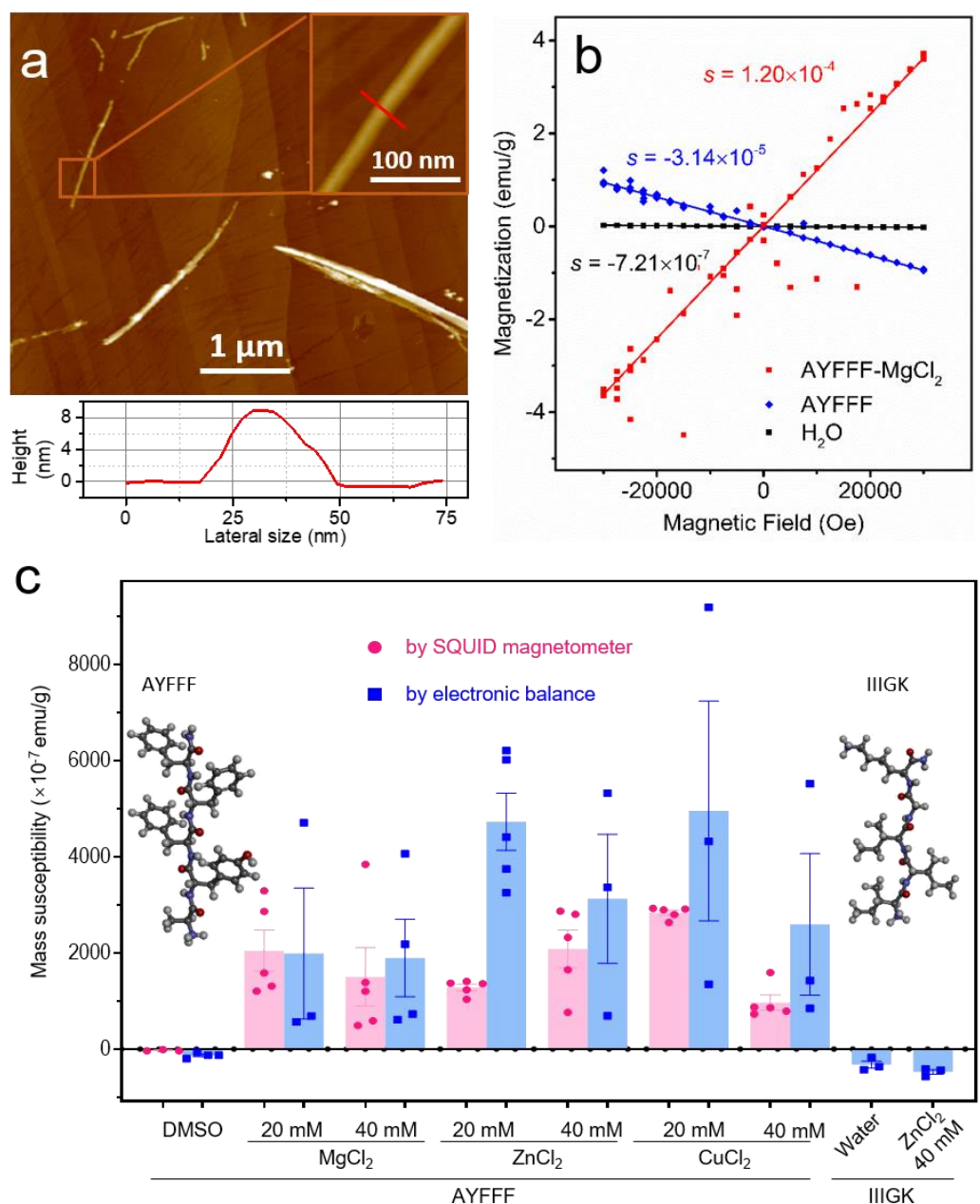

**Figure 1. Measurement of the average mass susceptibility (χ) of AYFFF peptide at different states in the chloride solution of divalent cations.** a) Atomic force microscopy (AFM) images of the self-assembled nanofibers of AYFFF aromatic peptides from the supernatant of the mixture with 40 mM $MgCl_2$ solution. The height profile of a typical nanofiber is shown below. b) Magnetization (M) versus magnetic

field (H) curves for typical AYFFF peptide assemblies dispersed in 40 mM $MgCl_2$ solution or pure water, using water as a reference. The symbol 's' represents the slope of the fitted lines. c) Average mass susceptibility ($\chi$) of AYFFF peptide at dissolved, or assembled states in different solutions, together with non-aromatic IIIGK assemblies as a control. The pink circles and blue squares display the experimental data collected by the SQUID magnetometer and the weighing method described in Ref. (31), respectively. Data are shown as mean ± SEM. The insets display the molecular structures of the AYFFF aromatic peptide and the IIIGK nonaromatic peptide. Note: both peptides are synthesized with the C-terminal amidation in order to increase their biological activities via generating a closer mimic of native proteins(32).

As mentioned in the Introduction, we presumed that the aromatic rings are associated with our observation of super strong paramagnetism. As a negative control, we performed measurements with assemblies of a non-aromatic peptide with a sequence of IIIGK. With the same methods as used for AYFFF peptides, we obtained the mass susceptibility of IIIGK assemblies in 40 mM $ZnCl_2$ solution, with an average value of $-3.2 \times 10^{-5}$ emu/g, which is not much different from that of IIIGK assemblies in pure water. This confirms that both aromatic rings and divalent cations are indispensable to the observed super strong paramagnetism.

Now we discuss the underlying physics of the observed super strong paramagnetism. Recently, we have reported the ferromagnetic properties of two-dimensional CaCl crystals on graphene, which is attributable to the monovalent calcium ions in CaCl crystals (22). We therefore hypothesize that there might be structures similar to CaCl with only one chloride ion adsorbed on one aromatic ring of AYFFF. We take the magnesium as an example to illustrate our hypothesis. For the convenience of description, we denote $MgCl_v$-$(H_2O)_n$ as the magnesium chloride hydrated with $n$ water molecules, with $v = 1$ or 2 representing the number of chloride ions; $MgCl_v$-$(H_2O)_n$@B is a $MgCl_v$-$(H_2O)_n$ cluster adsorbed on an aromatic ring. Here we used a benzene molecule as a typical aromatic ring in the computation, which is denoted as B. Using the density functional theory (DFT), we calculated the adsorption energy, denoted as $E_{ads}^{(v)}$, between a $MgCl_v$-$(H_2O)_n$ cluster and a benzene molecule as:

$$E_{ads}^{(v)} = E\left(MgCl_v\text{-}(H_2O)_n @ B\right) - E\left(MgCl_v\text{-}(H_2O)_n\right) - E(B) \quad (1),$$

where $E(MgCl_v$-$(H_2O)_n$@B), $E(MgCl_v$-$(H_2O)_n)$, and $E(B)$ are the total energies of a $MgCl_v$-$(H_2O)_n$ cluster adsorbed on a benzene molecule, a $MgCl_v$-$(H_2O)_n$ cluster, and a benzene molecule, respectively. The geometry of a $MgCl_v$-$(H_2O)_4$@B was fully optimized and the stable optimized geometries are displayed in Fig. 2a-b (other geometries with $n = 0-3$ are shown in Supplementary Fig. S4). Fig. 2c shows that the absolute value of the adsorption energy between a MgCl-$(H_2O)_n$ cluster and a benzene molecule increases significantly from 11.0 kcal/mol to 17.0 kcal/mol, as $n$ increases from 0 to 4. In contrast, the absolute value of the adsorption energy between $MgCl_2$-

($H_2O$)$_n$ and benzene decreases sharply from 18.1 kcal/mol to 8.1 kcal/mol, with the increase of $n$ from 0 to 4. Clearly, the absolute value (17.0 kcal/mol) of the adsorption energy of MgCl-($H_2O$)$_4$@B is significantly larger than that (8.1 kcal/mol) of $MgCl_2$-($H_2O$)$_4$@B, which indicates that the hydrated MgCl has a much higher occurrence probability than the hydrated $MgCl_2$ adsorbed on the benzene molecule in the solution. We note that the absolute value of the adsorption energy between the MgCl-($H_2O$)$_4$ cluster and the benzene molecule reaches ~28.3 $k_B$T for T = 300 K ($k_B$ is the Boltzmann's constant), which is about three times larger than the hydrogen-bond energy of water (~9 $k_B$T for T = 300 K (33)), implying that this adsorption is quite stable under room-temperature thermal fluctuations.

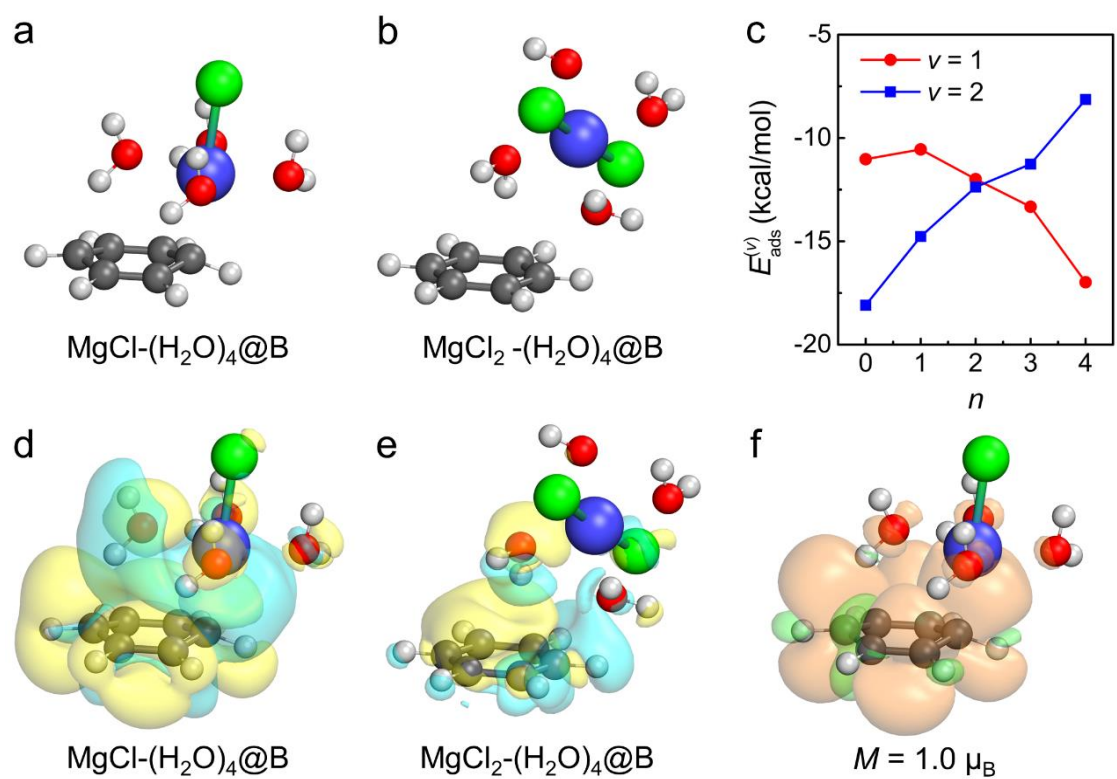

**Figure 2. Theoretical calculations for adsorption energies and magnetic properties of a hydrated magnesium chloride adsorbed on a benzene molecule.** a and b) Stable optimized geometry of clusters of (a) MgCl-($H_2O$)$_4$ and (b) $MgCl_2$-($H_2O$)$_4$ adsorbed on a benzene molecule (denoted as MgCl-($H_2O$)$_4$@B and $MgCl_2$-($H_2O$)$_4$@B, respectively). c) Adsorption energy of a MgCl$_v$-($H_2O$)$_n$ cluster adsorbed on a benzene molecule ($v$ = 1, solid circle; $v$ = 2, solid square) with $n$ = 0−4. d and e) Iso-surface of the charge-density difference resulting from adsorption of the (d) MgCl-($H_2O$)$_4$, and (e) $MgCl_2$-($H_2O$)$_4$ clusters on a benzene molecule. Iso-value of 0.0003 e/au$^3$. Yellow and cyan iso-surfaces represent charge accumulation and depletion in the space, respectively. f) Iso-surface spin density plot and the total magnetic moment (denoted as $M$) for the MgCl-($H_2O$)$_4$@B cluster. Orange and green iso-surfaces represent the positive and negative spin density of 0.0003 e/au$^3$, respectively. Spheres in green, blue, gray, red, and white represent chloride, magnesium, carbon, oxygen, and hydrogen, respectively.

We analyzed the charge density and molecular orbitals to understand the difference in the adsorption energies between MgCl-$(H_2O)_4$@B and MgCl$_2$-$(H_2O)_4$@B. The difference of charge density shows a clear charge re-distribution occurring at the Mg/benzene interfacial region in MgCl-$(H_2O)_4$@B (Fig. 2d), whereas no clear charge change is observed at the same region in MgCl$_2$-$(H_2O)_4$@B (Fig. 2e). In addition, molecular orbital analysis shows that there is a clear coupling between the lone-pair electrons of the Mg ion and the delocalized π states of the benzene molecule in MgCl-$(H_2O)_4$@B, but there is no such clear electronic coupling in the molecular orbitals of MgCl$_2$-$(H_2O)_4$@B (Supplementary Fig. S6). All these results revealed the electron transfer between the Mg ion and benzene in MgCl-$(H_2O)_4$@B, which enhanced the cation-π interaction between the Mg ion and the benzene as well as the adsorption energy. In contrast, in MgCl$_2$-$(H_2O)_4$@B, the cation-π interaction between the Mg ion and the benzene is strongly screened by the surrounding water molecules, resulting in a weak adsorption strength.

We further performed the spin-unrestricted DFT computations, which shows that the MgCl-$(H_2O)_4$ cluster adsorbed on the benzene molecule has a total magnetic moment of 1.0 $\mu_B$ per Mg and the spin density is mainly localized on the Mg ion and the benzene molecule (see Fig. 2f). It indicates that a MgCl-$(H_2O)_4$ cluster adsorbed on a benzene molecule with a monovalent magnesium ion can induce super strong paramagnetism.

Fluorescence and absorbance spectroscopy were both employed to show evidence of the cation-π interactions between the Mg ion and the aromatic ring structures in AYFFF peptide assemblies. The fluorescence spectrum of AYFFF supernatant excited at 275 nm has an emission peak at 303 nm, which is assigned to the aromatic ring structure in AYFFF component that easily generated the π-π* transition. Compared with the fluorescence intensity of AYFFF supernatant mixed with pure water, as shown in Fig. 3a, the fluorescence intensity of AYFFF supernatant containing 40 mM MgCl$_2$ markedly decreased, indicating that the conjugated double bonds of the aromatic rings in AYFFF are greatly affected in the MgCl$_2$ solution. Moreover, it was also observed that the UV absorption spectrum of AYFFF supernatant was affected by the cation-π interactions between magnesium ion and the aromatic ring of peptides (Supplementary Fig. S1). Comparison of the difference of spectra (AYFFF-MgCl$_2$ minus (AYFFF + MgCl$_2$)) with that of AYFFF reveals a weak positive band near 227 nm attributable to a Mg-ion-AYFFF interaction (26). These results further prove the existence of cation-π interactions between the aromatic ring of AYFFF and the Mg ion, which is consistent with our theoretical prediction.

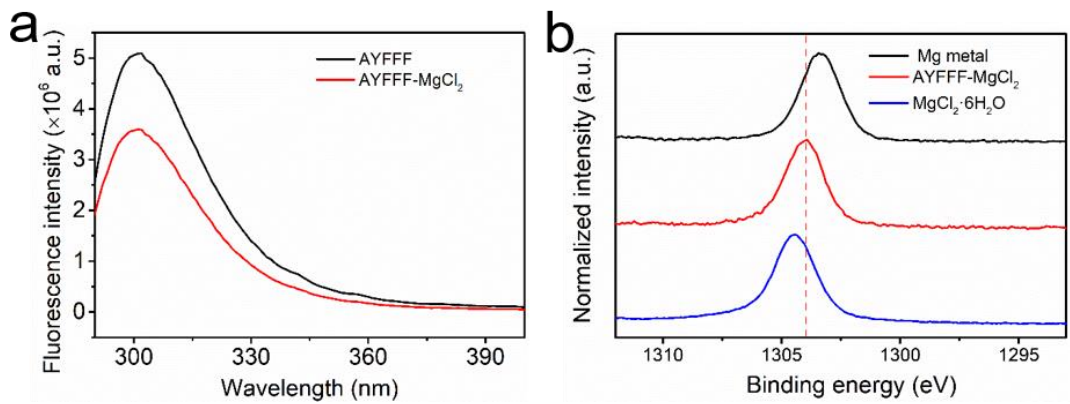

**Figure 3.** a) Fluorescence spectra of the mixed supernatant of AYFFF assemblies with MgCl$_2$ (red solid line) and with pure water (black solid line). b) X-ray photoelectron spectrometer (XPS) spectra of the Mg 1s peak of Mg metal, MgCl$_2$•6H$_2$O, and AYFFF-MgCl$_2$ nanofibers, respectively.

To confirm the abnormal valence-state change of cations in the AYFFF assemblies, we further performed X-ray photoelectron spectroscopy (XPS) experiments. In these experiments, the AYFFF assemblies were lyophilized after being collected from their salt solutions with an ultrafilter membrane (Mw cutoff: 300 kDa, Pall Corporation), in order to reduce the amount of ions from the solution before measurement. Fig. 3b shows that the Mg 1s peak is at 1303.9 eV for the AYFFF-MgCl$_2$ sample, in between the peak for regular MgCl$_2$•6H$_2$O at 1304.5 eV and for Mg metal at 1303.4 eV, indicating that the Mg ion presents a new valence state in the AYFFF assemblies, consistent with the DFT calculations.

The super strong paramagnetism of the AYFFF self-assemblies in the MgCl$_2$, ZnCl$_2$, and CuCl$_2$ solutions makes it possible to remotely manipulate aromatic ring-enriched materials using external magnetic fields. As shown in Fig. 4a, lots of large AYFFF assemblies precipitated to the bottom of the cuvette due to gravity, leaving the supernatant of small assemblies in 40 mM MgCl$_2$ at the first 2 minutes. When a magnet (~ 0.5 T) was close to the dispersion, it is clear that more and more peptide assemblies were attracted towards the magnet and finally accumulated along the left wall of the cuvette. In contrast, such phenomenon was not observable for AYFFF assemblies dispersed in pure water (Fig. 4b).

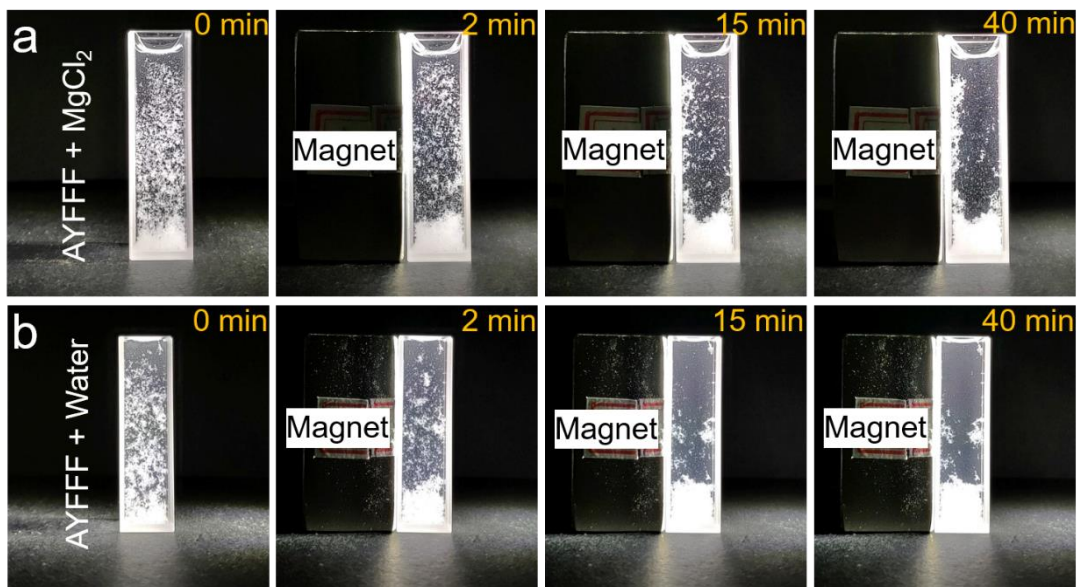

**Figure 4.** Snapshots of AYFFF assemblies dispersed in (a) 40 mM $MgCl_2$ or (b) pure water at different times as marked in photos.

We have also performed this long-range magnetic control experiment using the cellulose powder with a diameter of ~50 μm. We note that IIIGK assemblies are too thin to be seen, and the cellulose powder is visible and does not contain any aromatic rings as IIIGK. As expected, the magnetic field has negligible effect on the distribution of the cellulose powder, because there is no cation-π interaction between salts and the cellulose powder (Supplementary Fig. S2)

Finally, we would like to point out that the concentrations of Fe, Co, and Ni in the AYFFF assemblies and their salt solutions are negligibly small, only at the ppb level (Supplementary Table 1). In addition, the contribution of salt solutions themselves to the mass susceptibility of AYFFF assemblies has been subtracted as background during the data analysis. These further demonstrate that the super strong paramagnetism does not come from those traditional ferromagnetic materials.

We attribute the observation of super strong paramagnetism to the cations adsorbed on aromatic rings which display non-divalent behavior and thus possess magnetic moments. Considering that most of the aromatic rings inside the fibers might be free of cations so that they still contribute strong diamagnetism as reported in Ref. (31), we believe that AYFFF assemblies with much stronger paramagnetism, and even with the mass susceptibility of ferromagnetism, can be found in much thinner nanofibers or hollow nanotubes, in which a much higher proportion of aromatic rings have cations adsorbed. It should be mentioned that the AYFFF assemblies have relatively large variations in the diameter, which is the major reason why the mass susceptibilities for different samples can be quite different, i.e., a thinner nanofiber usually corresponds to a larger mass susceptibility, and vice versa (more below).

Our results are also inspirational for material fabrication and manipulation of aromatic ring-based paramagnetic biomaterials by cation addition. However, at present, we cannot separate the thinner nanofibers of AYFFF from the thicker fibers and separate the salts from the solution. Further work is required along this direction.

**Conclusion**

In summary, the mass susceptibilities obtained by both the SQUID and electrical balance experiments clearly demonstrated that the assemblies of an aromatic ring-enriched peptide in the chloride solution of various divalent metals display super strong paramagnetism. We attribute this phenomenon to existence of the magnetic moments on the cations adsorbed on aromatic rings in the AYFFF assemblies through hydrated cation-π interactions, where the adsorbed cations display non-divalent behavior with unpaired electron spins. Our results indicate the existence of super strong paramagnetism or even potential ferromagnetism in the aromatic ring-enriched biomolecules when there are enough cations of divalent metals adsorbed. Thus, our findings not only make an important step forward towards the understanding of the magnetism of biomolecules and re-thinking the origin of magnetoreception in living organisms, but also stimulate novel magnetic effects and magnetic controls on aromatic ring-enriched biomolecules and drugs, including their aggregation, dynamics, delivery and reactions, in living organisms as well as biomaterial fabrication and manipulation.